\documentclass[final,5p,times,twocolumn]{elsarticle}

\usepackage[hidelinks]{hyperref}

\usepackage[inline]{enumitem}
\usepackage{graphicx}
\usepackage[separate-uncertainty=true]{siunitx}
\usepackage{amsmath}   
\usepackage{mhchem}
\def\pmbanner{{\hrule height 1 pt}\vskip35pt{NIMA POST-PROCESS BANNER TO BE REMOVED AFTER FINAL ACCEPTANCE}\vskip35pt{\hrule height 4pt}\vskip20pt}
\begin{document}

\begin{frontmatter}

%% Note: \pmbanner before the actual title
\title{\pmbanner  Design and optimization of a hadronic calorimeter based on micropattern gaseous detectors for a future experiment at the Muon Collider}

\author[1]{Antonello Pellecchia\corref{cor1}}
    \ead{antonello.pellecchia@ba.infn.it}
\author[1,2]{Marco Buonsante}
\author[3]{Maryna Borysova}
\author[1,2]{Anna Colaleo}
\author[1]{Maria Teresa Camerlingo}
\author[1]{Luigi Longo}
\author[4]{Mauro Iodice}
\author[1]{Marcello Maggi}
\author[3]{Luca Moleri}
\author[1,2]{Raffaella Radogna}
\author[5]{Givi Sekhniaidze}
\author[1,2]{Federica Maria Simone}
\author[1,2]{Anna Stamerra}
\author[1,2]{Rosamaria Venditti}
\author[1]{Piet Verwilligen}
\author[3]{Darina Zavazieva}
\author[1,2]{Angela Zaza}

\cortext[cor1]{Corresponding author}

\affiliation[1]{
    organization={Istituto nazionale di fisica nucleare - Sezione di Bari}, 
    addressline={Via Amendola 173},
    city={Bari}, 
    country={Italy}
}
\affiliation[2]{
    organization={Università degli studi di Bari Aldo Moro}, 
    addressline={Piazza Umberto I},
    city={Bari}, 
    country={Italy}
}
\affiliation[3]{
    organization={Weizmann institute of science}, 
    addressline={Herzl St 234},
    city={Rehovot},
    country={Israel}
}
\affiliation[4]{
    organization={Istituto nazionale di fisica nucleare - Sezione di Roma 3}, 
    addressline={Via della Vasca Navale, 84},
    city={Roma},
    country={Italy}
}
\affiliation[5]{
    organization={Istituto nazionale di fisica nucleare - Sezione di Napoli}, 
    addressline={Via Cintia},
    city={Napoli},
    country={Italy}
}

\begin{abstract}
    Micro-pattern gaseous detectors (MPGDs) are a promising readout technology for hadronic calorimeters (HCAL) thanks to their good space resolution, longevity and rate capability.
    We describe the development of a HCAL based on MPGDs for an experiment at the proposed Muon Collider.
    The design of a semi-digital MPGD-HCAL is shown and its performance is calculated with Monte Carlo simulations with high-energy pions, showing an energy resolution down to 8\% for \SI{80}{\giga\eV} pions.
    We also present the performance of twelve MPGD prototypes with different technologies (MicroMegas, µ-RWELL and RPWELL) assembled and operated in test beam first with high-energy muons and later with pions in a hadronic calorimeter prototype of $\sim1\,\lambda_\text{I}$ length;
    the detectors have a good response uniformity (lower than 17\%) and space resolution and their performance in the calorimeter shows very good agreement with the Monte Carlo shower calculation.
\end{abstract}


\end{frontmatter}

%text of the article

\section{Introduction}
\label{sec:intro}

Calorimeters at future Higgs factories will require excellent energy resolutions to allow discriminating W and Z boson hadronic decays.
The proposed Muon Collider \cite{mucoll} requires a jet energy resolution of 3\%, which can be achieved by a hadronic calorimeter (HCAL) with resolution 60\%$/\sqrt{E}$;
to reach this performance target in the Muon Collider environment, chacterized by the presence of an asynchronous beam-induced-background (BIB), the hadronic calorimeter must have high readout granularity (with a readout element size between 1 and $3\times\SI{3}{\centi\m\squared}$) and a time resolution per cluster of the order of few \SI{}{\nano\s}.
A sampling hadronic calorimeter with micro-pattern gaseous detectors (MPGD) as active layers is expected to meet this reslution target.
Furthermore, MPGDs generally feature high radiation hardness, being able to withstand integrated charges of several \SI{}{\coulomb/\centi\m\squared}, and compared to other gaseous detector technologies, such as RPCs, have a better rate capability and space resolution, thus allowing better performance in presence of BIB.
The latest years have seen a rapid development of MPGDs using resistive elements -- such as resistive MicroMegas \cite{micromegas} and the µ-RWELL \cite{urwell} --, that compared to traditional MPGDs exibit better stability to discharge damage and are featured in several proposals of future collider systems \cite{urwell_idea} for tracking and timing layers \cite{picosec_chiara}.

%% Use \subsection commands to start a subsection.
\section{Calorimeter simulation}
\label{sec:design}

The performance of a MPGD-HCAL has been studied with a Geant4 \cite{geant} simulation.
The simulated detector geometry is made of alternating layers of \SI{2}{\centi\m}-thick iron absorbers and \SI{5}{\milli\m}-thick gaps of argon as readout MPGDs;
the nuclear interation length in this configuration is $\lambda_\text{I}=\SI{26}{\centi\m}$.
The size of each readout cell is $1\times\SI{1}{\centi\m\squared}$.

\begin{figure}[t]
    \centering
    \includegraphics[width=\linewidth]{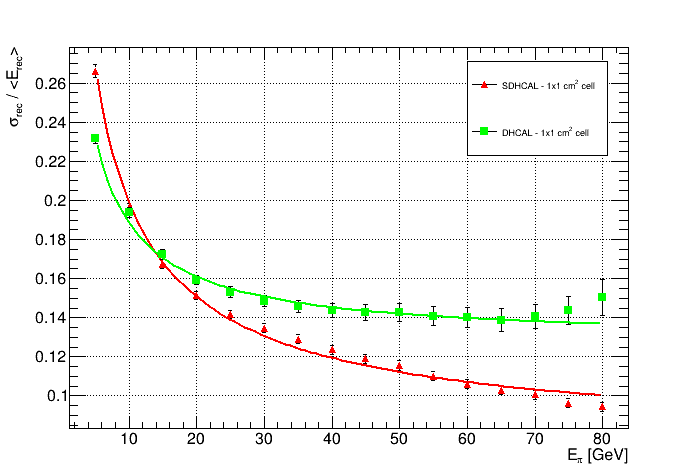}
    \caption{Simulated energy resolution of a $10\,\lambda_\text{I}$ MPGD calorimeter with digital and semi-digital readout.}\label{sim_energy}
\end{figure}

Results with a pion gun up to an energy of \SI{80}{\giga\eV} show that longitudinal shower containment is ensured with a length of $10\,\lambda_\text{I}$, while the transversal containment requires a width of $3\,\lambda_\text{I}$.
Using a semi-digital readout (SDHCAL \cite{sdhcal}), an energy resolution down to 8\% is found for an \SI{80}{\giga\eV} pion, while the resolution for digital readout (DHCAL) saturates at 14\% (Fig.\,\ref{sim_energy}).

\section{Performance of a calorimeter prototype}
\label{sec:prototype}

A prototype of MPGD-HCAL of length $1\,\lambda_\text{I}$ has been assembled and tested with two goals:
\begin{enumerate*}
    \item comparing the performance of different MPGD technologies and their suitability for calorimetry; and
    \item measuring the calorimeter energy resolution and comparing it to the Monte Carlo simulation results.
\end{enumerate*}

Twelve prototypes of readout detectors have been assembled: 7 µ-RWELL, 4 resistive MicroMegas and 1 RPWELL \cite{darina};
all detectors had an active area of $20\times\SI{20}{\centi\m\squared}$, a drift gap of thickness \SI{6}{\milli\m} and a readout made of 384 pads of $1\times\SI{1}{\centi\m\squared}$ size.

\subsection{Performance of MPGDs with MIPs}

A beam telescope made of the twelve MPGD prototypes has been tested with \SI{80}{\giga\eV}/c muons at the CERN North Area in 2023 to measure the detector efficiency and space resolution.
The detectors were operated with the state-of-the art gas mixture for each technology (\ce{Ar}:\ce{CO2}:\ce{C4H10} 93\%:5\%:2\% for the MicroMegas and RPWELL, \ce{Ar}:\ce{CO2}:\ce{CF4} 45\%:15\%:40\% for the µ-RWELL);
the front-end electronics was made of the APV25 \cite{apv25} ASIC\footnote{An analog front-end chip providing charge, arrival time and signal shape for each of the 128 readout channels.} read out by the RD51 scalable readout system (SRS) \cite{srs}.

The space resolution has been measured for each detector using as reference tracks the segments reconstructed in turn by all the other detectors in the telescope;
all detectors were found to have a good space resolution smaller than the readout pad pitch.
The response uniformity (measured as the relative sigma of the distribution of all the readout hits matching a muon track) is shown for seven of the detectors under test in Tab.\,\ref{uniformity}.
All MicroMegas and µ-RWELL prototypes have an excellent response uniformity of at most 16\%, with on average a better uniformity for the former compared to the latter;
more detailed studies are ongoing on the larger response uniformity of the µ-RWELL.
The RPWELL prototype also has a good response uniformity of \SI{22.6+-4.7}\%, larger than MicroMegas and µ-RWELL due to its larger amplification region.

\begin{table}[h]
\centering
\begin{tabular}{c | c}
    Detector ID & Uniformity (\%) \\ \hline
    MicroMegas Rm & \SI{12.3+-0.8}{} \\
    MicroMegas Na & \SI{11.6+-0.8}{} \\
    MicroMegas Ba & \SI{8.0+-0.5}{} \\
    RPWELL        & \SI{22.6+-4.7}{} \\
    µ-RWELL Na    & \SI{11.3+-1.0}{} \\
    µ-RWELL Fr1   & \SI{16.2+-1.7}{} \\
    µ-RWELL Fr2   & \SI{16.3+-1.1}{} \\
\end{tabular}
    \caption{Response uniformity of seven tested MPGD prototypes, calculated as the sigma -- extracted by a gaussian fit -- of the distribution of the charges of the muon hits.}\label{uniformity}
\end{table}

\subsection{Performance of calorimeter prototype with pions}

A calorimeter prototype of length $\sim1\,\lambda_\text{I}$ -- enough to ensure containment of a \SI{10}{\giga\eV} pion -- has been assembled and tested with pion beams in the CERN East Area in 2023.
The test beam setup was made of 8 layers of iron absorbers and readout MPGDs;
the thickness of the absorbers was \SI{4}{\centi\m} for the first two layers -- to anticipate the starting point of the pion shower -- and \SI{2}{\centi\m} for the subsequent ones.

\begin{figure}[t]
    \centering
    \includegraphics[width=\linewidth]{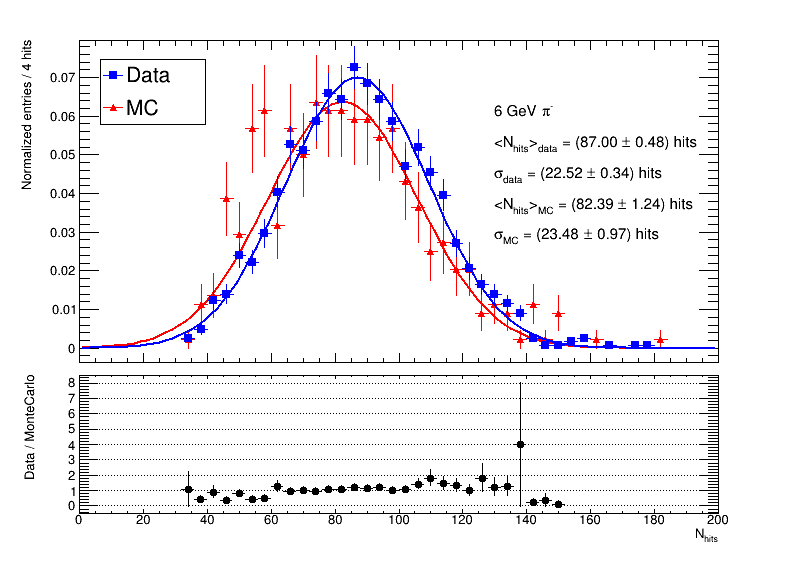}
    \caption{Distribution of the total number of hits in the calorimeter prototype tested at the CERN PS East Area for a \SI{6}{\giga\eV} pion, compared with the results of a Monte Carlo simulation performed with the same geometry.}\label{ps_data_mc}
\end{figure}

The calorimeter response has been measured with pions between 1 and \SI{11}{\giga\eV};
the data were analyzed in a digital readout approach, i.e. counting the total number of hits over a fixed threshold in the calorimeter event by event.
To compare the test beam data with the Monte Carlo simulation of the prototype, the efficiency of each readout layer\footnote{Including both intrinsic detector efficiency and electronics effects, such as dead readout channels.} was accounted for in the simulation.
The data to Monte Carlo comparison (Fig.\,\ref{ps_data_mc} for a \SI{6}{\giga\eV} pion) shows good agreement for the entire pion energy range;
a refinement of the analysis to obtain a comparison to the semi-digital readout approach and a full energy calibration is ongoing.

\section{Conclusions and outlook}

Sampling hadronic calorimeters with MPGD active layers are a promising technology for environments with high asyncronous background for experiments at future colliders such as the Muon Collider.
The performance of a digital and semi-digital calorimeter tower has been studied in simulations with pions up to \SI{80}{\giga\eV}, showing a promising energy resolution down to 8\%.
Twelve $20\times\SI{20}{\centi\m\squared}$ MPGD prototypes using three different technologies (resistive MicroMegas, µ-RWELL and RPWELL) have been assembled and tested with muons in test beam showing good space resolution and efficiency uniformity.
Finally, a calorimeter prototype made of eight layers of iron absorbers and MPGDs has been built and tested with pions of 1 to \SI{11}{\giga\eV} using analog readout electronics;
the measured calorimeter response shows good agreement between data and Monte Carlo in the digital readout approach.

The development of the MPGD-HCAL will continue with testing the response uniformity of larger-area MPGDs and with the integration of semi-digital front-end electronics to ensure the scalability to a larger number of layers.

\bibliographystyle{elsarticle-num-names} 
\bibliography{main}

\end{document}